\documentclass[12pt]{article}

\usepackage{amsfonts}
\usepackage{amsmath}
\usepackage{url}
\usepackage{natbib}

\newtheorem{thm}{Theorem}[section]
\newtheorem{rem}{Remark}[section]
\newtheorem{lem}{Lemma}[section]
\newtheorem{prob}{Problem}[section]
\newtheorem{prop}{Proposition}[section]

\newcommand{\A}{\mathcal{A}}
\newcommand{\Po}{\mathbb{P}}
\newcommand{\Q}{\mathbb{Q}}

\begin{document}
\title{Optimal Investment and Risk Control Problem for an Insurer: Expected Utility Maximization}
\author{Bin Zou\thanks{ Email: bzou@ualberta.ca. Phone: (+1)780-716-7987.} and Abel Cadenillas\thanks{Corresponding author: Central Academic Building 639, University of Alberta, Edmonton T6G 2G1 Canada. Email: abel@ualberta.ca. Phone: (+1)780-492-0572.}\\Department of Mathematical and Statistical Sciences\\University of Alberta}

\maketitle

\begin{abstract}
Motivated by the AIG bailout case in the financial crisis of 2007-2008, we consider an insurer who wants to maximize the expected utility of the terminal wealth by selecting optimal investment and risk control strategies. The insurer's risk process is modelled by a jump-diffusion process and is negatively correlated with the capital gains in the financial market. We obtain explicit solution to optimal strategies for various utility functions.
\end{abstract}

\section{Introduction}
\label{sec_intro}
The financial crisis of 2007-2008 caused a significant recession in global economy, considered by many economists to be the worst financial crisis since the Great Depression of the 1930s. It resulted in the threat of bankruptcy of large financial institutions, the bailout of banks, and downturns in stock markets around the world (See more on Wikipedia \url{http://en.wikipedia.org/wiki/Financial_crisis_of_2007-08}). American International Group, Inc. (AIG), once the largest insurance companies in the United States with a triple-A credit rating, collapsed within a few months in 2008. The stock price of AIG was traded at over \$50 per share in Februray, but plunged down to less than \$1 per share when AIG was on the brink of bankruptcy. The severity of AIG's liquidity crisis led to an initial rescue of \$85 billion and a total of \$182 billion bailout by the U.S. government, the largest government bailout in history. See \citet[Chapter 6]{stein}  for more statistical data of AIG during the past financial crisis and \cite{sjostrom} for detailed discussions on AIG bailout case. According to \citet[Chapter 6]{stein}, AIG made several major mistakes which together contributed to its sudden collapse. First, AIG underpriced the risk of writing Credit Default Swap (CDS) contracts since it ignored the negative correlation between its liabilities and the capital gains in the financial market. Second, AIG applied a problematic model for risk management and failed to estimate the impact of the risk on the company's capital structure.
To address these issues, we propose a jump-diffusion process to model AIG's risk (per policy risk) and consider optimal investment and risk control problem  for an insurer like AIG. So our research has two root in the literature: optimal consumption and investment problem and optimal reinsurance (risk control) problem.

\cite{merton} was the first to apply stochastic control theory to solve consumption and investment problem in continuous time. \cite{karatzas1} provided a rigorous analysis to Merton's problem. Later \cite{karatzas2} further generalized the results in an incomplete market. \cite{zhou} and \cite{sotomayor} improved the model by incorporating regime switching. They also obtained explicit solutions under the mean-variance criterion and the utility maximization criterion, respectively. \cite{moore} incorporated another random risk (which can be insured against by purchasing insurance policy) into Merton's framework and studied optimal consumption, investment and insurance problem for the first time. Following \cite{moore}, \cite{perera} revisited the same problem in a more general Levy market.
Along the same vein, many researchers added an uncontrollable risk process to Merton's model. They then consider a stochastic control problem for optimal investment strategy (without consumption mostly) under certain criteria. For instance, \cite{browne} modeled the risk process using a geometric Brownian motion and studied optimal investment problem under two different criteria: maximizing the expected exponential utility of the terminal wealth and minimizing the probability of ruin. \cite{wang} applied a jump-diffusion model for the risk process and considered optimal investment problem under the utility maximization criterion. For stochastic control theory with jumps and its applications to finance, please see two monographs, \cite{cont} and \cite{oksendal}.

In mathematics, there are two main tools for solving stochastic control problems. The first tool is dynamic programming and maximum principle, see, for instance, \cite{fleming2} and \cite{cadenillas}. The second tool is martingale approach, based on equivalent martingale measures and martingale representation theorems. The martingale approach and its application in continuous time finance was developed by \cite{harrison1}. Thereafter, martingale method has been applied to solve many important problems in finance. For example, option pricing problem in \cite{harrison2}, optimal consumption and investment problem in \cite{karatzas2}, optimal consumption, investment and insurance problem in \cite{perera} and optimal investment problem in \cite{wang}. In Section 3, we also apply martingale approach to solve our stochastic control problem.

The second root of our research is optimal reinsurance problem, which studies an insurer who wants to control the reinsurance payout for certain objectives. Reinsurance is an important tool for insurance companies to manage their risk exposure. The classical model for risk in the insurance literature is Cramer-Lundberg model, which uses a compound Poisson process to measure risk. The Cramer-Lungberg model was introduced by Lungberg in 1903 and then republished by Cramer in 1930s. Since the limiting process of a compound Poisson process is a diffusion process, see \cite{taksar}, recent research began to model risk by a diffusion process or a jump-diffusion process, see \cite{wang}.  \cite{hojgaard} assumed the reserve of an insurance company is governed by a diffusion process and consider the optimization criterion of maximizing the expected utility of running reserve up to the bankruptcy time. \cite{kaluszka} studied optimal reinsurance in discrete time under mean-variance criterion for both proportional reinsurance and step loss reinsurance. \cite{schmidli} considered both the Cramer-Lundberg model and the diffusion model for the risk process and obtained optimal proportional reinsurance policies under the criterion of minimizing the ruin probability. Recent generalizations in modeling for optimal reinsurance problem include incorporating regime switching, see \cite{zhuo}, and interest rate risk and inflation risk, see \cite{guan}.

Our model and optimization problem are different from the existing ones in the literature in several directions. Comparing with Merton's framework and its generalizations, we add a controllable jump-diffusion process into the model, which will be used to model the insurer's risk (per policy risk). We then regulate the risk for the insurer by controlling the number of policies. So our model is also different from the ones considered in optimal reinsurance problem and its variants, which control risk by purchasing reinsurance policies from another insurer. As suggested in  \citet[Chapter 6]{stein}, we assume there is a negative correlation between the financial market (capital gains) and the risk (liabilities) in our model. \citet[Chapter 6]{stein} considered a similar risk regulation problem as ours, but in his model, investment strategy is fixed and the risk is modelled by a geometric Brownian motion. To generalize Stein's work, we model the risk by a jump-diffusion process and allow the insurer to choose investment strategy continuously. \citet[Chapter 6]{stein} considered the problem only with logarithmic utility function, which can be easily solved using classic stochastic method. We obtain explicit solutions to optimal investment and risk control problem for various utility functions, including hyperbolic absolute risk aversion (HARA) utility function (logarithmic function and power function), constant absolute risk aversion (CARA) utility function (exponential function) and quadratic utility function.

The structure of this paper is organized as follows. We describe our model and formulate optimal investment and risk control problem in Section \ref{sec_model}. We obtain explicit solutions to optimal investment and risk control strategies for logarithmic utility function in Section \ref{sec_log}, power utility function in Section \ref{sec_power}, exponential utility function in Section \ref{sec_exp} and quadratic utility function in Section \ref{sec_quad}. We conclude our study in Section \ref{sec_conclusion}.

\section{The Model}
\label{sec_model}
In our model, there are two trading assets in the financial market, a riskless asset $P_0$ and a risky asset (mutual fund) $P_1$. On a filtered probability space $(\Omega, \mathcal{F}, \{\mathcal{F}_t\}_{t \ge 0}, \mathbb{P})$, the dynamics of $P_0$ and $P_1$ are given by
\begin{align*}
dP_0(t)&=r(t)P_0(t)dt,\\
dP_1(t)&=P_1(t)(\mu(t)dt+\sigma(t)dW^{(1)}(t)),
\end{align*}
where $r$, $\mu$ and $\sigma$ are positive bounded functions and $W^{(1)}$ is a standard Brownian motion. The initial conditions are $P_0(0)=1$ and $P_1(0)>0$.

For an insurer like AIG, its main liabilities come from writing insurance policies, and we denote the total outstanding number of policies (liabilities) at time $t$ by $L(t)$. In the actuarial industry, the premium is usually precalculated, which means insurance companies charge premium based on historical data and estimation models. For example, regarding auto insurance policies, insurance companies consider several main factors, such as the insured's demographic information, previous driving record, coverage needs, and the type of vehicle, then use an actuarial model to calculate the premium for the insured. Therefore, it is reasonable to assume premium per policy is a fixed constant for a certain type of insurance contracts and a given group of the insured. To simplify our analysis, we further assume the average premium per policy for the insurer is $p$, so the revenue from selling insurance policies over the time period of $(t,t+dt)$ is given by $pL(t)dt$.

A commonly used risk model for claims in the actuarial industry is compound Poisson model (Cram{\'e}r-Lundberg model), in which the claim (risk) per policy is given by $\sum\limits_{i=1}^{N(t)} Y_i$, where $\{Y_i\}$ is a series of independent and identical distributed random variables, and $N(t)$ is a Poisson process independent of $Y_i$. If the mean of $Y_i$ and the intensity of $N(t)$ is finite, then such compound Poisson process is a Levy process with finite Levy measure. According to  \cite{oksendal}, a Levy process can be decomposed into there components, a linear drift part, a Brownian motion part and a pure jump part. Based on this result, we assume the total risk is given by
\begin{equation*}
dR(t)=L(t) (adt+bd\bar{W}(t) + \gamma dN(t)), \, R(0)=0,
\end{equation*}
where $\bar{W}$ is a standard Brownian motion and $N$ is a Poisson process defined on the given filtered space, respectively. We assume $a, b, \gamma$ are all positive constants. \cite{stein} argues that one of the most serious mistakes AIG made was ignoring the negative correlations between its liabilities and the capital gains in the financial market. So we assume
\begin{equation*}
\bar{W}(t)=\rho W^{(1)}(t) + \sqrt{1-\rho^2} W^{(2)}(t),
\end{equation*}
where $\rho <0$ and $W^{(2)}$ is another standard Brownian motion, independent of $W^{(1)}$. We also assume the Poisson process $N$ has a constant intensity $\lambda$, and is independent of both $W^{(1)}$ and $W^{(2)}$.

At time $t$, an insurer (AIG) chooses $\tilde{\pi}(t)$, the dollar amount invested in the risky asset, and
total liabilities $L(t)$. For a strategy $\tilde{u}:=(\tilde{\pi},L)$, the corresponding wealth process (surplus process) $X^{\tilde{u}}$  is driven by the following SDE
\begin{equation} \label{wealth1}
\begin{split}
dX^{\tilde{u}}(t) &=  \left( r(t) X^{\tilde{u}}(t) + (\mu(t)-r(t))\tilde{\pi}(t) + (p-a)L(t) \right)dt - \gamma L(t) dN_t \\
&\quad+(\sigma(t) \tilde{\pi}(t) - \rho b L(t))dW^{(1)}(t) - b \sqrt{1-\rho^2}L(t) dW^{(2)}(t),
\end{split}
\end{equation}
with initial wealth $X^{\tilde{u}}(0)=x>0$.

Following \citet[Chapter 6]{stein}, we define the ratio of liabilities over surplus as $\kappa(t):=\frac{L(t)}{X(t)}$ (which is called debt ratio). We denote $\pi(t)$ as the proportion of wealth invested in the risky asset at time $t$. Then for a control $u(t):=(\pi(t),\kappa(t))$, we have $\tilde{u}(t)=X(t) u(t)$. We then rewrite SDE \eqref{wealth1} as
\begin{equation} \label{wealth2}
\begin{split}
\frac{dX^u(t)}{X^u(t-)}&= (r(t)+ (\mu(t)-r(t))\pi(t) + (p-a) \kappa(t))dt -\gamma \kappa(t)dN(t) \\
&\quad  + (\sigma(t) \pi(t) - b \rho \kappa(t)) dW^{(1)}(t)- b\sqrt{1-\rho^2} \kappa(t)dW^{(2)}(t),
\end{split}
\end{equation}
with $X^{u}(0)=x>0$.

\begin{rem}
In a financial market, it is universally acknowledged that extra uncertainty (risk) must be compensated by extra return. So in our model, we impose further conditions on the coefficients: $\mu(t)>r(t)\ge 0$ and $p>a>0$.
\end{rem}

We define the criterion function as
\begin{equation} \label{criterion1}
J(x;u)=E_x \left[U(X^u(T)) \right],
\end{equation}
where $E_x$ means conditional expectation under probability measure $\mathbb{P}$ with $X^u(0)=x$ and $T>0$ is the terminal time. Utility function $U$ is assumed to be a strictly increasing and concave function. The common choices for utility function in economics and finance are $U(x)=\ln(x)$, $U(x)=-\frac{1}{\alpha} e^{-\alpha x}$, where $\alpha>0$, and $U(x)=x^\alpha$, where $\alpha<1$ and $\alpha \neq 0$.

Expected utility maximization, as probably the most widely used optimization criterion in economics and finance, has been used in various investment/consumption and reinsurance problems. To name a few, for instance, \cite{merton}, \cite{karatzas2}, and \cite{wang}.

We denote $\A_x$ as the set of all admissible controls with initial wealth $X(0)=x$. Depending on the utility function, we choose either $u$ or $\tilde{u}$ to be our control and then formally define the admissible set $\A_x$. The value function is defined by
\begin{equation*}
V(x):=\sup_{u \in \A_x} J(x;u),
\end{equation*}
where $u$ will be changed accordingly if the control we choose is $\tilde{u}$.

We then formulate our stochastic control problem as follows.
\begin{prob}  \label{prob_opt}
Select an admissible control $u^*=(\pi^*,\kappa^*) \in \A_x$ (or $\tilde{u}^*=(\tilde{\pi}^*,L^*) \in \A_x$) that attains the value function $V(x)$.
The control $u^*$ (or $\tilde{u}^*$) is called an optimal control or an optimal policy.
\end{prob}

\section{The Analysis for $U(x)=\ln(x),\,x>0$}
\label{sec_log}

We first consider Problem \ref{prob_opt} when the utility function is given by $U(x)=\ln(x),\,x>0$,
which belongs to the class of hyperbolic absolute risk aversion (HARA) utility functions.

We choose $u$ as control and denote $\mathcal{A}_1$ as the set of all admissible controls when $U(x)=\ln(x)$. For every $u \in \mathcal{A}_1$,
$\{u(t)\}_{0\le t \le T}$ is progressively measurable with respect to the filtration $\{\mathcal{F}_t\}_{0\le t \le T}$ and $\forall \, t \in [0,T]$, satisfies the following conditions
\begin{equation*}
\int_0^t \pi(s)^2 ds <\infty,\text{ and }\int_0^t \kappa(s)^2 ds <\infty,\,\kappa(t) \ge 0.
\end{equation*}
Furthermore, to avoid the possibility of bankruptcy at jumps, we assume $\kappa(t) < \frac{1}{\gamma}$ if $u \in \mathcal{A}_1$.

Notice that $\forall \, u \in \A_1$, SDE \eqref{wealth2} satisfies the linear growth condition and Lipschitz continuity condition, so by Theorem 1.19 in \cite{oksendal}, there exists a unique solution $X^u$ such that
\[ E [|X^u(t)|^2] < \infty \quad \text{for all }t \in [0,T].\]

\begin{prop}
\label{prop_positive_wealth}

Under optimal control $u^*$ of Problem \ref{prob_opt}, the associated optimal wealth $X^{u^*}(t)$ is strictly positive for all $t \in [0,T]$.
\end{prop}
\emph{Proof.}
Notice that $u_0:=(\pi \equiv 0, \kappa \equiv 0) \in \A_1$ is an admissible control, and under the control $u_0$, the wealth $X^{u_0}$ is given by
\[X^{u_0}(t)=x \, e^ {\int_0^t r(s)ds} >0, \, \forall \, t \in [0,T].\]
So we can conclude, under optimal control $u^*$, $X^{u^*}(t) \ge X^{u_0}(t)>0$ for all $t \in [0,T]$. To see this conclusion, assume to the contrary that for some
$t' \in [0,T]$, $ x':=X^{u_0}(t') > X^{u^*}(t')$. We define a new control $u'$ by
\begin{equation*}
u'(t):= \begin{cases}
u_0(t), & t\in [0,t'];\\
u^*(t), & t\in (t',T].
\end{cases}
\end{equation*}
By definition, $u' \in \A_1$ and $X^{u'}(t')=X^{u_0}(t') > X^{u^*}(t')$. Recall the strong Markov property of $X$, we obtain $X^{u'}(T) > X^{u^*}(T)$,
and hence
\[J(x,u') > J(x,u^*),\]
a contraction to the fact that $u^*$ is optimal control to Problem \ref{prob_opt}. \hfill $\Box$

\begin{rem}
Thanks to Proposition \ref{prop_positive_wealth}, we do not need to deal with the bankruptcy time, as discussed in \cite{sotomayor}, in our analysis.
\end{rem}

We then apply two methods to solve Problem \ref{prob_opt} when the admissible set is $\mathcal{A}_1$.

\subsection{Method 1: Optimization Method in Calculus}
\label{subsec_log_control}
Under the logarithmic utility assumption, we can apply the classic optimization method in calculus to solve Problem \ref{prob_opt}. For more details on using this method to solve stochastic control problems, please see \citet[Chapter 4,5,6]{stein}.

Applying Ito's formula to $\ln(X_t)$, we obtain
\begin{align*}
\ln \frac{X^u_t}{X_0} &= \int_0^t \Big(r_s +(\mu_s-r_s)\pi_s +(p-a)\kappa_s -\frac{1}{2}\sigma_s^2 \pi_s^2 +\rho b\sigma_s \pi_s \kappa_s \\
&\quad -\frac{1}{2}b^2 \kappa_s^2 + \lambda \ln(1-\gamma \kappa_s) \Big)ds + \int_0^t \left(\sigma_s\pi_s - b\rho \kappa_s \right)dW^{(1)}_s \\
&\quad -\int_0^t b\sqrt{1-\rho^2}\kappa_s dW^{(2)}_s + \int_0^t \ln(1-\gamma \kappa_s) dM_s,
\end{align*}
where $M_t:=N_t -\lambda t$ is the compensated Poisson process of $N$ and is a martingale under $\mathbb{P}$.

For any given $u \in \mathcal{A}_1$, we have
\begin{align*}
\int_0^t \left(\sigma_s\pi_s - b\rho \kappa_s \right)^2 ds &\le K_1 \int_0^t \pi_s^2 ds +K_2 \int_0^t \kappa_s^2 ds<\infty,\\
\int_0^t b^2 (1-\rho^2)\kappa_s^2 ds &\le K_3 \int_0^t \kappa_s^2 ds<\infty,
\end{align*}
for some positive constants $K_i$, $i=1,2,3$.

Therefore, we obtain
\[E_x \left[ \int_0^t \left(\sigma_s\pi_s - b\rho \kappa_s \right)dW^{(1)}_s\right] =E_x \left[\int_0^t b\sqrt{1-\rho^2}\kappa_s dW^{(2)}_s \right]=0.\]

Since $\kappa$ is a bounded predictable process, so is $\ln(1-\gamma \kappa)$ and then implies the stochastic integral $\int_0^t \ln(1-\gamma \kappa_s) dM_s$ is again a $\mathbb{P}$-martingale with the initial value being $0$. So we obtain
\[ E_x \left[\int_0^t \ln(1-\gamma \kappa_s) dM_s \right]=0.\]

The above analysis yields
\[ E_x \left[\ln \frac{X^u_t}{X^u_0} \right] = E_x \left[\int_0^T f(\pi(t),\kappa(t))dt \right],\]
where $f(\pi(t),\kappa(t)):=r(t)+(\mu(t)-r(t))\pi(t) +(p-a)\kappa(t) -\frac{1}{2}\sigma(t)^2 \pi(t)^2 +\rho b\sigma(t) \pi(t) \kappa(t)-\frac{1}{2}b^2 \kappa(t)^2 + \lambda \ln(1-\gamma \kappa(t))$.

Hence we obtain the optimization condition as follows
\[ u^*(t)=\arg \sup _{u\in \mathcal{A}_1} J(x;u)=\arg\sup_{u\in \mathcal{A}_1} f(\pi(t),\kappa(t)).\]

The first-order condition is then given by
\begin{equation} \label{firstordercond}
\begin{split}
(\mu(t)-r(t)) - \sigma^2(t) \pi^*(t) + \rho b \sigma(t) \kappa^*(t) &=0,\\
(p-a) + \rho b \sigma(t) \pi^*(t) - b^2 \kappa^*(t) - \frac{\lambda \gamma}{1-\gamma \kappa^*(t)} &=0.
\end{split}
\end{equation}

The candidate investment proportion in the risky asset $\pi^*$ will be
\begin{equation} \label{optpi}
\pi^*(t)=\frac{\mu(t)-r(t)+\rho b \sigma(t) \kappa^*(t)}{\sigma^2(t)},
\end{equation}
where $\kappa^*(t)$ is the solution to the following quadratic equation
\begin{equation} \label{optka}
A (\kappa^*(t))^2 - B(t) \, \kappa^*(t) + C(t)=0,
\end{equation}
with $A:=b^2(1-\rho^2) \gamma$, $B(t):=b^2(1-\rho^2)+\gamma(p-a +\rho\, b \, \dfrac{\mu(t)-r(t)}{\sigma(t)})$ and $C(t):=p-a +\rho\, b \, \dfrac{\mu(t)-r(t)}{\sigma(t)} -\lambda \gamma$.

It is easy to check that
\[ \Delta(t):=B^2(t)-4A \, C(t)=(b^2(1-\rho^2)-\gamma(C(t)+\lambda \gamma))^2+4\lambda b^2 (1-\rho^2) \gamma^2>0.\]
So the quadratic system \eqref{optka} has two solutions and one is given by
\[ \kappa_+(t)=\frac{B(t)+\sqrt{\Delta(t)}}{2A}>\frac{1}{\gamma},\]
which is not included in the admissible set $\A_1$.

To ensure the existence of a non-negative $\kappa^* \in [0,\frac{1}{\gamma})$, we impose a technical condition $\min_{t\in[0,T]} C(t) >0$, which is equivalent to
\begin{equation} \label{techcondition}
p-a +\rho\, b \, \dfrac{\mu(t)-r(t)}{\sigma(t)} > \lambda \gamma, \, \forall \, t \in [0,T].
\end{equation}

When the technical condition \eqref{techcondition} holds, we have
\begin{equation} \label{optkappa}
\kappa^*(t)=\kappa_-(t):=\frac{B(t)-\sqrt{\Delta(t)}}{2A}.
\end{equation}

Notice that a sufficient condition for a regular interior maximizer and the first-order condition to be hold is
\begin{equation} \label{optcondition}
f_{\pi\pi}<0, \, f_{\kappa\kappa}<0, \text{ and } f_{\pi\pi}f_{\kappa\kappa} - f^2_{\pi\kappa}>0.
\end{equation}
We then calculate those partial derivatives and verify that the above condition \eqref{optcondition} is satisfied.
\begin{align*}
f_{\pi\pi} &=-\sigma^2(t)<0, \\
f_{\kappa\kappa} &= -b^2 - \frac{\lambda \gamma^2 \kappa(t)}{(1-\gamma \kappa(t))^2}<0,\\
f_{\pi\pi}f_{\kappa\kappa} - f^2_{\pi\kappa} &= (1-\rho^2) b^2 \sigma^2(t) + \frac{\lambda \gamma^2 \sigma^2(t) \kappa(t)}{(1-\gamma \kappa(t))^2}>0.
\end{align*}

\begin{thm} \label{logthm}
When $U(y)=\ln(y)$, and the technical condition \eqref{techcondition} holds, $u^*(t)=(\pi^*(t),\kappa^*(t))$, where $\pi^*(t)$ and $\kappa^*(t)$ are given by \eqref{optpi} and \eqref{optkappa}, respectively, is optimal control to
Problem \ref{prob_opt} with the admissible set $\mathcal{A}_1$.
\end{thm}
\emph{Proof.} $\forall \, u=(\pi,\kappa) \in \mathcal{A}_1$, since $u^*$ defined above is the maximizer of $f$, we have
\[f(\pi^*(t),\kappa^*(t)) \ge f(\pi(t),\kappa(t)), \, \forall \, t \in[0,T],\]
and then
\[\int_0^T f(\pi^*(t),\kappa^*(t))dt \ge \int_0^T f(\pi(t),\kappa(t))dt,\]
which implies $J(x,u^*) \ge J(x,u)$. Due to the arbitrariness of $u$, we obtain $J(x,u^*) \ge V(x)$.

To complete the proof, we then verify that $u^*$ is admissible.

Since $\frac{C(t)}{A}>0$ and $\kappa_+(t)>0$, we get $\kappa^*(t)=\kappa_-(t)>0$.

To show $\kappa^*(t) < \frac{1}{\gamma}$, it is equivalent to show
\[ \Delta(t) > (\gamma (C(t)+\lambda \gamma) - b^2(1-\rho)^2)^2,\]
which is always satisfied if we recall the definition of $\Delta(t)$.

So we have $0\le \kappa^*(t) < \frac{1}{\gamma}$, which in turn implies
\[\int_0^t (\kappa^*(s))^2 ds  < \infty, \forall \, t\in [0,T].\]

From our assumption, $\mu(t)$, $r(t)$ and $\sigma(t)$ are all positive and bounded functions, for all $t \in [0,T]$, we obtain
\[ \int_0^t (\pi^*(s))^2 ds \le K_4 t + K_5 \int_0^t (\kappa^*(s))^2 ds  < \infty,\]
for some positive constants $K_4$ and $K_5$.

Therefore $u^*$ defined above is an admissible control and then is optimal control to Problem \ref{prob_opt}. \hfill $\Box$

\subsection{Method 2: Martingale Method}
\label{subsec_log_martingale}
In this subsection, we apply the martingale method to solve Problem \ref{prob_opt}. To begin with, we give two important Lemmas, which are Proposition 2.1 and Lemma 2.1 in \cite{wang}, respectively. Lemma \ref{optlemma} gives the condition optimal control must satisfy. Lemma \ref{lem_rep} is a generalized version of martingale representation theorem.
Please consult \cite{wang} and \citet[Chapter 9]{cont} for details.

\begin{lem} \label{optlemma}
If there exists a control $u^* \, (\text{or }\tilde{u}^*) \in \mathcal{A}$ such that
\begin{equation} \label{martingalecondition}
E\left[ U'(X^{u^*}(T)) \, X^{u}(T)\right] \text{ is constant over all admissible controls},
\end{equation}
then $u^*$ (or $\tilde{u}^*$) is optimal control to Problem \ref{prob_opt}.
\end{lem}

\begin{lem}
\label{lem_rep}
For any $\Po$-martingale $Z$, there exists predictable processes $\theta=(\theta_1,\theta_2,\theta_3)$ such that
\[ Z_t=Z_0 + \int_0^t \theta_1(s)dW_s^{(1)}+\int_0^t \theta_2(s)dW_s^{(2)}+\int_0^t \theta_3(s)dM_s,\]
for all $t \in [0,T]$.
\end{lem}

We then find optimal control to Problem \ref{prob_opt} through the three steps.\\
\vspace{1ex}\\
\textbf{Step 1.} We conjecture the candidates for optimal strategies $\pi^*$ and $\kappa^*$.

Define
\begin{equation} \label{logzetadef}
Z_T:=\frac{(X_T^{u^*})^{-1}}{E[(X_T^{u^*})^{-1}]}, \text{ and }Z_{\eta}:=E[Z_t|\mathcal{F}_{\eta}]
\end{equation}
for any stopping time $\eta \le T$ almost surely.  Recall Proposition \ref{prop_positive_wealth}, the process $Z$ is a strictly positive (square-integrable) martingale under $\Po$ with $E(Z_t)=1$, for all $t \in [0,T]$. Then we can define a new measure $\Q$ by $\dfrac{d \Q}{d \mathbb{P}}:=Z_T$.

From SDE \eqref{wealth1}, we have
\begin{align*}
X^u_t=X^{\tilde{u}}_t &= xe^{rt} + \int_0^t e^{r(t-s)} ((\mu_s-r_s)\tilde{\pi}_s + (p-a)L_s) ds - \int_0^t e^{r(t-s)} \gamma L_s dN_s \\
&\quad +\int_0^t e^{r(t-s)}(\sigma_s \tilde{\pi}_s - \rho b L_s)dW_s^{(1)} - \int_0^t e^{r(t-s)} b \sqrt{1-\rho^2}L_s dW_s^{(2)}.
\end{align*}

Using the above expression of $X$ and Lemma \ref{optlemma}, for all admissible controls, we have
\begin{equation} \label{logcondition}
\begin{split}
E_{\mathbb{Q}}\Big[ \int_0^t e^{-rs} \Big( &((\mu_s-r_s)\tilde{\pi}_s + (p-a)L_s) ds + (\sigma_s \tilde{\pi}_s - \rho b L_s)dW_s^{(1)}  \\
& -  b \sqrt{1-\rho^2}L_s dW_s^{(2)}- \gamma L_s dN_s \Big)\Big] \text{ is constant.}
\end{split}
\end{equation}

We define
\[K_t := \int_0^t \frac{1}{Z_{s-}} dZ_t, \, t \in [0,T].\]
Since $Z$ is a $\mathbb{P}$-martingale, so is $K$.

By Lemma \ref{lem_rep}, there exist predictable processes $(\theta_1, \theta_2, \theta_3)$ such that(One can consult \cite{wang} for measurability and integrability conditions $\theta$ should satisfy.)
\begin{equation*}
dK_t=\theta_1(t)dW_t^{(1)} + \theta_2(t) dW_t^{(2)} + \theta_3(t) dM(t).
\end{equation*}

Then by Doleans-Dade exponential formula, we have
\begin{equation} \label{zeta}
\begin{split}
Z_t=\exp\Big\{ &\int_0^t (\theta_1(s)dW_s^{(1)}+\theta_2(s)dW_s^{(2)}+\ln(1+\theta_3(s))dN_s)\\
&-\frac{1}{2}\int_0^t (\theta_1^2(s)+\theta_2^2(s)-2\lambda \theta_3(s))ds \Big\}.
\end{split}
\end{equation}

By Girsanov's Theorem, $W^{(i)}(t)-\int_0^t \theta_i(s)ds$, $i=1,2$, is a Brownian Motion under $\mathbb{Q}$ and $N(t)-\int_0^t \lambda (1+\theta_3(s))ds$ is a martingale under $\mathbb{Q}$.

For any stopping time $\eta \le T$, we choose $\tilde{\pi}(t)=1_{t \le \eta}$ and $L(t)=0$, which is apparently an admissible control. By substituting this control into \eqref{logcondition}, we obtain
\[ E_{\mathbb{Q}} \left[ \int_0^\eta e^{-rs}(\mu_s-r_s)ds + \int_0^\eta e^{-rs} \sigma_s dW_s^{(1)} \right] \text{ is constant over all } \eta \le T,\]
which implies
\begin{equation}
\int_0^t e^{-rs}(\mu_s-r_s)ds + \int_0^t e^{-rs} \sigma_s dW_s^{(1)} \text{ is a $\mathbb{Q}$-martingale.}
\end{equation}

Therefore, $\theta_1$ must satisfy the equation
\begin{equation*}
\mu(t)-r(t) + \sigma(t) \theta_1(t) =0,
\end{equation*}
or equivalently,
\begin{equation} \label{log1}
\theta_1(t)=-\frac{\mu(t)-r(t)}{ \sigma(t) }, \, t\in [0,T].
\end{equation}

Next we choose $\tilde{\pi}(t)=0$ and $L(t)=1_{t \le \eta}$. By following a similar argument as above, we have
\[\int_0^t e^{-rs} ((p-a)ds-\rho b dW_s^{(1)} - b \sqrt{1-\rho^2} dW_s^{(2)} - \gamma dN_s )\text{ is a $\mathbb{Q}$-martingale},\]
which in turn yields
\begin{equation*}
p-a - \rho b \theta_1(t) - b \sqrt{1-\rho^2} \theta_2(t) - \lambda \gamma (1+\theta_3(t))=0, \, t\in[0,T].
\end{equation*}

By \eqref{log1}, we can rewrite the above equation as
\begin{equation} \label{log2}
p-a + \rho b \, \frac{\mu(t)-r(t)}{ \sigma(t) } - b \sqrt{1-\rho^2} \, \theta_2(t) - \lambda \gamma (1+\theta_3(t))=0, \, t\in[0,T].
\end{equation}

\begin{rem} \label{martingaleremark}
Notice that the above analysis holds for all utility functions except that the definition of $Z$ in \eqref{logzetadef} changes accordingly. More importantly, we emphasize that the conditions \eqref{log1} and \eqref{log2} are satisfied for all utility functions, although $\theta_2$ and $\theta_3$ will be different for different utility functions. We shall use the conclusion in this remark when applying martingale approach to solve Problem \ref{prob_opt} for different utility functions thereafter.
\end{rem}

From SDE \eqref{wealth2}, we can solve to get $(X^{u^*}_T)^{-1}$
\begin{equation} \label{logzeta}
\begin{split}
(X^{u^*}_T)^{-1} = x^{-1} \exp \Big\{ &-\int_0^T f(\pi^*_t,\kappa^*_t) dt -\int_0^T (\sigma_t \pi_t^* - \rho b \kappa^*_t)dW_t^{(1)} \\
&+ \int_0^T b\sqrt{1-\rho^2}\kappa^*_t dW_t^{(2)} - \int_0^T \ln(1-\gamma \kappa^*_t) dM_t\Big\}.
\end{split}
\end{equation}

By comparing the $dW^{(1)}$, $dW^{(2)}$ and $dN$ terms in \eqref{zeta} and \eqref{logzeta}, we obtain
\begin{equation} \label{comparecond1}
\begin{split}
\theta_1(t)&=-(\sigma(t) \pi^*(t) - \rho b \kappa^*(t),\\
\theta_2(t)&=b\sqrt{1-\rho^2} \kappa^*(t),\\
\ln(1+\theta_3(t))&=-\ln(1-\gamma \kappa^*(t)).
\end{split}
\end{equation}

By plugging \eqref{comparecond1} into \eqref{log1} and \eqref{log2}, we obtain the same system \eqref{firstordercond} as in Method 1. So we find the same optimal strategies $\pi^*$ and $\kappa^*$, which are given by \eqref{optpi} and \eqref{optkappa}, respectively.\\
\vspace{1ex}\\
\textbf{Step 2.} For $\theta_i$ given in \eqref{comparecond1} and $u^*=(\pi^*,\kappa^*)$ defined by \eqref{optpi} and \eqref{optkappa}, we verify that $Z_T$ defined by \eqref{zeta} is consistent with its definition.

We first rewrite \eqref{logzeta} as
\[  \frac{1}{X_T^{u^*}} = I_T H_T,\]
where
\begin{align*}
I_T:&=\frac{1}{x} \exp\left\{ \int_0^T (-f(\pi^*_s,\kappa^*_s) + \lambda \ln(1-\gamma \kappa^*_s))ds \right\},\\
H_T:&=\exp\Big\{ -\int_0^T (\sigma_s \pi_s^* - \rho b \kappa^*_s)dW_s^{(1)}+ \int_0^T b\sqrt{1-\rho^2}\kappa^*_sdW_s^{(2)} \\
 &\quad\quad\quad\;\; - \int_0^T \ln(1-\gamma \kappa^*_s)dN_s \Big\}.
\end{align*}

By substituting \eqref{comparecond1} back into \eqref{zeta}, we obtain
\[Z_T=J_T H_T,\]
where
\[J_T:=\exp\left\{\int_0^T (-\frac{1}{2} \sigma_s^2 (\pi^*_s)^2 + \rho b \sigma_s \pi^*_s \kappa_s^* -\frac{1}{2}b^2(\kappa_s^*)^2+\lambda (\frac{1}{1-\gamma \kappa_s^*}-1))ds \right\}\]
is constant.

By definition, we know $Z$ is a $\mathbb{P}$-martingale and $E[Z_T]=1$, and then
\[ E[H_T]=\frac{1}{J_T}.\]

Therefore, we obtain
\[ Z_T=\frac{(X^{u^*}_T)^{-1}}{E \left[(X^{u^*}_T)^{-1}\right]}=\frac{I_T H_T}{I_T E[H_T]}=\frac{H_T}{J_T^{-1}}=J_TH_T,\]
which shows $Z$ given by \eqref{zeta} with $\theta_i$ provided by \eqref{comparecond1} is the same as the definition: $Z_T=\frac{(X^{u^*}_T)^{-1}}{E \left[(X^{u^*}_T)^{-1}\right]}$.\\
\vspace{1ex}\\
\textbf{Step 3.} We verify $\pi^*$ and $\kappa^*$, given by \eqref{optpi} and \eqref{optkappa}, respectively, are indeed optimal strategies. Equivalently, we verify
the condition \eqref{martingalecondition} is satisfied for $u^*=(\pi^*, \kappa^*)$.

For any $u\in \A_1$, we define a new process $Y^u$ as follows
\begin{align*}
Y_t^u:=&\int_0^t e^{-rs}X_s^u \left( (\mu_s-r_s)\pi_s + (p-a)\kappa_s \right) ds - \int_0^t e^{-rs}X_s^u \gamma \kappa dN_s \\
 &\; +\int_0^t e^{-rs}X_s^u \left( (\sigma_s\pi_s -\rho b \kappa_s)dW_s^{(1)} - b \sqrt{1-\rho^2} \kappa_s dW_s^{(2)} \right)\\
=& \int_0^t e^{-rs}X_s^u \left(p-a+\rho b \frac{\mu_s-r_s}{\sigma_s}-b^2(1-\rho^2) \kappa_s^* - \frac{\lambda \gamma}{1-\gamma \kappa_s^*} \right) ds\\
&\quad + \text{ local $\mathbb{Q}$-martingale}.
\end{align*}

Due to the first-order condition \eqref{firstordercond}, the above $ds$ term will be $0$, and then $Y^u$ is a local $\mathbb{Q}$-martingale.

Since $u^*$ is deterministic and bounded, $Z$ is a square-integrable martingale under $\mathbb{P}$, which implies $E[(Z_T)^2] < \infty$ or equivalently, $Z \in L^2(\mathcal{F})$. Furthermore, for any $u \in \mathcal{A}_1$, we have $X^u\in L^2(\mathcal{F})$, so is $Y^u$. Therefore, we have
\[E_\mathbb{Q} \left[ \sup_{0\le t \le T} |Y_t^u| \right] \le \sqrt{E[(Z_T)^2]} \sqrt{E\left[ \sup_{0\le t \le T} |Y_t^u|^2\right]} < \infty,\]
which enables us to conclude that the family
\[\{ Y_{\eta}^u: \text{ stopping time } \eta \le T\}\text{ is uniformly integrable under $\mathbb{Q}$}.\]
Hence $Y^u$ is indeed a martingale under $\mathbb{Q}$ with $E_{\mathbb{Q}} [Y_t^u]=0$ for any $u \in \mathcal{A}_1$. This result verifies the condition \eqref{martingalecondition} is satisfied.

Therefore, Lemma \ref{optlemma} together with the above three steps lead to Theorem \ref{logthm}. \hfill $\Box$

\section{The Analysis for $U(y)=y^{\alpha},\,0<\alpha<1$}
\label{sec_power}
The second utility function we consider is power function, which also belongs to HARA class. Here, we choose $\mathcal{A}_1$ as the admissible set for Problem \ref{prob_opt}.

Since $U'(X_T^{u^*})=\alpha (X_T^{u^*})^{\alpha -1}$, we define $Z$ as
\begin{equation} \label{powerzetadef}
Z_T:=\frac{\left(X_T^{u^*} \right)^{\alpha-1}}{E\left[\left(X_T^{u^*} \right)^{\alpha-1} \right]}, \text{ and } Z_{\eta}:=E[Z_T|\mathcal{F}_{\eta}],
\end{equation}
where $\eta$ is a stopping time and $\eta \le T$ almost surely. With the help of $Z$, we can define a new probability measure $\mathbb{Q}$ as $\dfrac{d \mathbb{Q}}{d \mathbb{P}}=Z_T$.

From SDE \eqref{wealth2}, we obtain
\begin{equation} \label{powerzeta}
\begin{split}
\left(X_T^{u^*} \right)^{\alpha-1}=\text{constant} \, \cdot \, &\exp\bigg\{ \int_0^T (\alpha-1)\Big( (\sigma_t \pi_t^* - \rho b \kappa_t^*)dW_t^{(1)} \\
&-b\sqrt{1-\rho^2} \kappa_t^* dW_t^{(2)}+\ln(1-\gamma \kappa_t^*)dN_t\Big)\bigg\}.
\end{split}
\end{equation}

Thanks to Remark \ref{martingaleremark}, $Z_T$ also bears the expression \eqref{zeta}.
So by comparing the terms of $dW^{(1)}$, $dW^{(2)}$ and $dN$ in \eqref{zeta} and \eqref{powerzeta}, we obtain
\begin{equation} \label{comparecond2}
\begin{split}
\theta_1(t) &= (\alpha-1) (\sigma(t)\pi^*(t) - \rho b \kappa^*(t)),\\
\theta_2(t) &= -b(\alpha-1) \sqrt{1-\rho^2} \kappa^*(t),\\
\ln(1+\theta_3(t)) &= (\alpha-1) \ln(1-\gamma \kappa^*(t)).
\end{split}
\end{equation}

Substituting $\theta_1$ in \eqref{comparecond2} into \eqref{log1}, we obtain optimal proportion $\pi^*$ of investment in the risky asset
\begin{equation} \label{powerpi}
\pi^*(t)=-\frac{\mu(t)-r(t)}{(\alpha-1) \sigma^2(t)} + \frac{\rho \, b}{\sigma(t)} \kappa^*(t),
\end{equation}
with $\kappa^*$ will be determined below by \eqref{powerkappa}.

Due to Remark \ref{martingaleremark}, $\theta_2$ and $\theta_3$ defined above should satisfy the equation \eqref{log2}. We then plug \eqref{comparecond2} into \eqref{log2}, and obtain
\begin{equation} \label{powerkappa}
p-a+\rho b \, \frac{\mu(t)-r(t)}{\sigma(t)}+(\alpha-1)b^2(1-\rho^2)\kappa^*(t)-\lambda \gamma (1-\gamma \kappa^*(t))^{\alpha-1}=0.
\end{equation}

Define
\begin{align*}
\phi(t):&=1-\gamma \kappa^*(t),\\
B_1:&=\dfrac{(\alpha-1) b^2 (1-\rho^2)}{\lambda \gamma^2},\\
C_1(t):&=-\dfrac{1}{\lambda \gamma} \bigg[p-a+\rho b \dfrac{\mu(t)-r(t)}{\sigma(t)} + \dfrac{(\alpha-1) b^2 (1-\rho^2)}{\gamma} \bigg].
\end{align*}
Then the equation \eqref{powerkappa} for optimal debt ratio $\kappa^*$ can be rewritten as
\begin{equation} \label{powercond}
\left( \phi(t) \right)^{\alpha-1} + B_1 \, \phi(t) +C_1(t)=0.
\end{equation}

\begin{lem} \label{powerlemma}
If the condition \eqref{techcondition} holds, there exists a unique solution $\phi(t) \in (0,1)$ to the equation \eqref{powercond}, and then exists a unique solution $\kappa^*(t) \in (0,\frac{1}{\gamma})$ to the equation \eqref{powerkappa}.
\end{lem}
\emph{Proof.} Define $h(x):=x^{\alpha-1}+B_1 \, x + C_1(t)$. It is easy to check $h'(x)=(\alpha-1)x^{\alpha-2} + B_1 <0$ since $\alpha-1<0$ and $B_1<0$. Besides, $\lim_{x \to 0^+} h(x) = +\infty$. Due to the technical condition \eqref{techcondition}, we have
\[h(1)=1-\dfrac{1}{\lambda \gamma} \left(p-a+\rho b \, \frac{\mu(t)-r(t)}{\sigma(t)} \right)<0.\]
Hence, there exists a unique solution in $(0,1)$ to the equation \eqref{powercond} for all $t \in [0,T]$. Recall the definition of $\phi$, if $\phi \in (0,1)$, then $\kappa^* \in (0,\frac{1}{\gamma})$, and so the equation \eqref{powerkappa} also bears a (unique) solution in $(0,\frac{1}{\gamma})$. \hfill $\Box$

\begin{thm} \label{powerthm}
When $U(y)=y^{\alpha}$, $0<\alpha<1$, and the technical condition \eqref{techcondition} holds, $u^*(t)=(\pi^*(t),\kappa^*(t))$, with $\pi^*$ and $\kappa^*$ given by \eqref{powerpi} and \eqref{powerkappa}, respectively, is optimal control to Problem \ref{prob_opt} with the admissible set $\mathcal{A}_1$.
\end{thm}
\emph{Proof.} Because of Lemma \ref{powerlemma}, $\pi^*$ and $\kappa^*$ given by \eqref{powerpi} and \eqref{powerkappa} are well-defined if the condition \eqref{techcondition} is satisfied. By following Steps 2 and 3 as in Section \ref{sec_log}, we can verify that the condition \eqref{martingalecondition} holds for the above defined $u^*=(\pi^*,\kappa^*)$. Then it remains to show that $u^*$ is admissible.

By Lemma \ref{powerlemma}, we have $\kappa^*(t) \in (0,\frac{1}{\gamma})$, and then the square integrability condition for $\kappa^*$ follows. Recall \eqref{powerpi} and $\mu,r,\sigma$ are all bounded, so $\pi^*$ is also square-integrable.
Therefore, $u^* \in \mathcal{A}_1$ and then $u^*$ is optimal control to Problem \ref{prob_opt}. \hfill $\Box$

\begin{rem}
We notice that the analysis in this section still holds when the utility function is given by $U(x)=c-x^{\alpha}$, $\alpha<0$. Since $U'(x)=-\alpha x^{\alpha-1}$, we define $Z_T$ to be the same as \eqref{powerzetadef}. Furthermore, when $\alpha<0$, we have $\alpha-1<0$,  so all the results in Lemma \ref{powerlemma} and Theorem \ref{powerthm} follow as well.
\end{rem}

\section{The Analysis for $U(x)=-\frac{1}{\alpha} e^{-\alpha x},\,\alpha>0$}
\label{sec_exp}
In this section, we consider Problem \ref{prob_opt} for exponential utility function, which is of constant absolute risk aversion (CARA) class. We define the admissible set $\mathcal{A}_2$ as follows: for any admissible control $\tilde{u}=(\tilde{\pi},L) \in \mathcal{A}_2$, $\{\tilde{u}\}_{0\le t \le T}$ is progressively measurable with respect to the filtration $\{\mathcal{F}\}_{0 \le t \le T}$, and satisfies the integrability conditions
\begin{equation*}
E \left[\int_0^t (\tilde{\pi}(s))^2 ds \right] <\infty, \, E \left[\int_0^t (L(s))^2 ds \right]<\infty,
\end{equation*}
and $L(t) \ge 0$, $\forall \, t \in [0,T]$.

In what follows, we apply the martingale approach to find optimal control to Problem \ref{prob_opt} with the admissible set as $\mathcal{A}=\mathcal{A}_2$.

By Lemma \ref{optlemma}, optimal control $\tilde{u}^*$ should satisfy the following condition
\begin{equation} \label{expcond1}
E \left[ \exp\{-\alpha X_T^{\tilde{u}^*}\} \, X_T^{\tilde{u}} \right] \text{ is constant for all } u \in \mathcal{A}_2.
\end{equation}

So we define Rando-Nikodym process by
\begin{equation}
Z_T:=\frac{e^{-\alpha X_T^{u^*}}}{E[e^{-\alpha X_T^{u^*}}]}, \text{ and }Z_{\eta}:=E[Z_T|\mathcal{F}_{\eta}],
\end{equation}
for any stopping time $\eta \le T$, and a new probability measure $\mathbb{Q}$ by
$\frac{d \mathbb{Q}}{d \mathbb{P}}=Z_T$.

Since $Z$ is a martingale under $\mathbb{P}$,  exists progressively measurable process $\theta_i$, $i=1,2,3$ such that $Z$ is in the form of \eqref{zeta}.

From SDE \eqref{wealth1}, we can calculate
\begin{equation} \label{expzeta}
\begin{split}
\exp\left(-\alpha X_T^{\tilde{u}^*}\right)=\text{constant} \cdot \exp\bigg\{& -\int_0^T \alpha e^{r(T-t)} \Big( (\sigma_t \tilde{\pi}^*_t - \rho b L^*_t)dW_t^{(1)} \\
&-b\sqrt{1-\rho^2} L^*_t dW_t^{(2)} - \gamma L^*_t dN_t\Big) \bigg\}.
\end{split}
\end{equation}

Comparing \eqref{zeta} and \eqref{expzeta} gives
\begin{equation} \label{comparecond3}
\begin{split}
\theta_1(t) &= - \alpha e^{r(T-t)} (\sigma(t) \tilde{\pi}^*(t) - \rho b L^*(t)),\\
\theta_2(t) &=  \alpha e^{r(T-t)} b\sqrt{1-\rho^2} L^*(t),\\
\ln(1+\theta_3(t)) &= \alpha \gamma e^{r(T-t)} L^*(t).
\end{split}
\end{equation}

By \eqref{log1}, we have
\begin{equation} \label{exppi}
\tilde{\pi}^*(t)=e^{-r(T-t)} \frac{\mu(t)-r(t)}{\alpha \sigma(t)^2} + \frac{\rho b}{\sigma(t)} L^*(t).
\end{equation}

Substituting \eqref{comparecond3} into \eqref{log2}, we obtain
\begin{equation} \label{expkappa}
\lambda \gamma \, e^{A_3(t) L^*(t)} + B_3(t) \, L^*(t) - C_3(t)=0,
\end{equation}
with $A_3,\, B_3$ and $C_3$ defined by
\begin{align*}
A_3(t):&=\alpha \, \gamma \, e^{r(T-t)},\\
B_3(t):&=\alpha e^{r(T-t)}b^2(1-\rho^2),\\
C_3(t):&=p-a+\rho b \, \frac{\mu(t)-r(t)}{\sigma(t)}.
\end{align*}

\begin{lem} \label{explem}
If the condition \eqref{techcondition} holds, then there exists a (unique) positive solution to the equation \eqref{expkappa}.
\end{lem}
\emph{Proof.}
We define $\tilde{h}(x):=\lambda \, \gamma  \, e^{A_3(t)x} + B_3(t) \, x - C_3(t)$. Since $\tilde{h}'(x)=\lambda \, \gamma  \, A_3(t) e^{A_3(t)x}  + B_3(t)$ and $A_3(t)>0$, $B_3(t)>0$ for all $t \in [0,T]$, we have $\tilde{h}'(x)>0$. Because the condition \eqref{techcondition} holds, we obtain $\tilde{h}(0)=\lambda \gamma -C_3(t)<0$ for all $t \in [0,T]$. Besides, $C_3(t)$ is a bounded function on $[0,t]$ and then has a finite maximum, which implies $\tilde{h}(x)>0$ when $x$ is large enough. Therefore, as a continuous and strictly increasing function, $\tilde{h}(x)$ has a (unique) positive zero point.\hfill $\Box$.

\begin{thm}
When $U(y)=-\frac{1}{\alpha}e^{-\alpha y}$, $\alpha>0$, $\tilde{u}^*(t)=(\tilde{\pi}^*(t),L^*(t))$, where $\tilde{\pi}^*$ and $L^*$ are defined by \eqref{exppi} and \eqref{expkappa}, respectively, is optimal control to Problem \ref{prob_opt} with the admissible set $\mathcal{A}_2$ .
\end{thm}
\emph{Proof.} Please refer to Theorem \ref{powerthm} for proof. \hfill $\Box$

\section{The Analysis for $U(x)=x - \dfrac{\alpha}{2} x^2,\,\alpha>0$}
\label{sec_quad}
As pointed in \cite{wang}, to find a mean-variance portfolio strategy is equivalent to maximize the expected utility for a quadratic function. So in this section, we consider a quadratic utility function, and solve Problem \ref{prob_opt} with admissible set $\mathcal{A}=\mathcal{A}_2$. Notice that quadratic utility function is not strictly increasing for all $x$, but rather has a maximum point at $x=\frac{1}{\alpha}$. This means if the investor's wealth is greater than the maximum point, he/she will experience a decreasing utility as wealth keeps rising. Such result is consistent with the famous efficient frontier theory (discovered by \cite{markowitz}).

Since $U'(y)=1-\alpha y$, our objective is to find $\tilde{u}^* \in \mathcal{A}_2$ such that
\begin{equation} \label{quadcond1}
E[(1-\alpha X^{\tilde{u}^*}_T) X_T^{\tilde{u}}] \text{ is constant for all } u \in \mathcal{A}_2.
\end{equation}

Define $Z_T:=1-\alpha X^{\tilde{u}^*}_T$ and $Z_t:=E[Z_T|\mathcal{F}_t]$. Since $\tilde{u}^* \in \mathcal{A}_2$, $Z$ is a square-integrable martingale under $\mathbb{P}$, and therefore there exists progressively measurable processes $\theta_i$, $i=1,2,3$ such that
\begin{equation*}
dZ(t)=\theta_1(t)dW^{(1)}(t)+\theta_2(t)dW^{(2)}(t)+\theta_3(t)dM(t).
\end{equation*}

Define process $\tilde{Y}^{\tilde{u}}$ by
\begin{align*}
\tilde{Y}^{\tilde{u}}(t):=\int_0^t e^{-rs}\big[ &((\mu_s-r_s)\tilde{\pi}_s+(p-a)L_s)ds +(\sigma_s \tilde{\pi}_s-\rho b L_s)dW_s^{(1)} \\
&- b \sqrt{1-\rho^2}L_s dW_s^{(2)} - \gamma L_s dN_s \big].
\end{align*}
Then we can write $X^{\tilde{u}}_t$ as $X^{\tilde{u}}(t)=e^{rt} (x+\tilde{Y}^{\tilde{u}}(t))$ and obtain a sufficient condition for \eqref{quadcond1}
\[ \{ \tilde{Y}^{\tilde{u}}(t)Z(t)\}_{t\in [0,T]} \text{ is a martingale under measure } \mathbb{P}.\]

By Ito's formula, we have
\begin{align*}
d\tilde{Y}^{\tilde{u}}_tZ_t &=\tilde{Y}^{\tilde{u}}_{t-}dZ_t + Z_{t-}d\tilde{Y}^{\tilde{u}}_t + d[\tilde{Y}^{\tilde{u}},Z](t) \\
&=\tilde{Y}^{\tilde{u}}_{t-}dZ_t+Z_{t-}e^{-rt} \big( (\mu_t-r_t) \tilde{\pi}_t+(p-a)L_t \big)dt \\
&\quad + Z_{t-} e^{-rt} (\sigma_t \tilde{\pi}_t - \rho b L_t)dW_t^{(1)} -Z_{t-}e^{-rt} b \sqrt{1-\rho^2} L_t dW_t^{(2)}\\
&\quad-Z_{t-}e^{-rt}\gamma L_t dN_t +\theta_1(t)e^{-rt}(\sigma_t \tilde{\pi}_t - \rho b L_t)dt \\
&\quad-\theta_2(t)e^{-rt} b \sqrt{1-\rho^2} L_tdt -\theta_3(t)e^{-rt}\gamma L_t dN_t.
\end{align*}

Then a necessary condition for $\tilde{Y}^{\tilde{u}}Z$ to be a $\mathbb{P}$-martingale is
\begin{equation*}
\begin{split}
&\quad Z_{t-} ((\mu_t-r_t) \tilde{\pi}_t+(p-a)L_t -\lambda \gamma L_t) + \theta_1(t)(\sigma_t \tilde{\pi}_t - \rho b L_t) \\
&-\theta_2(t) b \sqrt{1-\rho^2} L_t - \theta_3(t) \lambda \gamma L_t=0.
\end{split}
\end{equation*}
By considering two admissible controls $(\tilde{\pi}=1,L=0)$ and $(\tilde{\pi}=0,L=1)$, we obtain
\begin{align}
&Z_{t-} (\mu(t)-r(t)) + \sigma(t) \theta_1(t)=0 \; \Rightarrow \theta_1(t)=-\frac{\mu(t)-r(t)}{\sigma(t)} Z_{t-}. \label{quadtheta1}\\
&Z_{t-}(p-a-\lambda \gamma) -\rho b \theta_1(t) - b\sqrt{1-\rho^2} \theta_2(t) - \lambda \gamma \theta_3(t)=0. \label{quadtheta2}
\end{align}

Define $P(t):=\exp\{ \int_0^t \xi(s)ds \}$, $t\in [0,T]$, where $\xi$ is a deterministic function and will be determined later. Applying Ito's formula to $P_tZ_t$ gives
\begin{align*}
P_T Z_T &= Z_0 +\int_0^T P_t dZ_t + \int_0^T Z_{t-}dP_t \\
&= Z_0+ \int_0^T Z_{t-} \xi_t P_t dt - \int_0^T \frac{\mu_t-r_t}{\sigma_t} Z_{t-}P_t dW_t^{(1)} \\
&\quad+ \int_0^T P_t \theta_2(t)dW_t^{(2)}+\int_0^T P_t \theta_3(t)dN_t-\int_0^T \lambda P_t \theta_3(t)  dt.
\end{align*}

Recall the definition of $Z_T$, we obtain $X^{\tilde{u}^*}_T=\frac{1-Z_T}{\alpha}=\frac{1}{\alpha} - \frac{P_T Z_T}{\alpha P_T} $ and
\begin{equation} \label{quadwealth}
\begin{split}
X^{\tilde{u}^*}_T &= \frac{1}{\alpha} - \frac{Z_0}{\alpha P_T} - \frac{1}{\alpha P_T} \int_0^T Z_{t-} \xi_t P_tdt\\
&\quad +\frac{1}{\alpha P_T}  \int_0^T \frac{\mu_t-r_t}{\sigma_t} Z_{t-}P_t dW_t^{(1)} - \frac{1}{\alpha P_T} \int_0^T P_t \theta_2(t)dW_t^{(2)} \\
&\quad - \frac{1}{\alpha P_T} \int_0^T P_t \theta_3(t)dN_t +  \frac{1}{\alpha P_T}\int_0^T \lambda P_t \theta_3(t)  dt.
\end{split}
\end{equation}

By substituting optimal control $\tilde{u}^*$ into SDE \eqref{wealth1}, we solve to get
\begin{align}
X^{\tilde{u}^*}_T &=xe^{rT} + \int_0^T e^{r(T-t)} \big( ((\mu_s-r_s)\tilde{\pi}^*_s+(p-a)L^*_s \big)ds- \int_0^T e^{r(T-t)}\gamma L^*_t dN_t \notag\\
&\quad +\int_0^T e^{r(T-t)} \left( (\sigma_t \tilde{\pi}^*_t-\rho b L^*_t)dW_t^{(1)} - b \sqrt{1-\rho^2}L^*_t dW_t^{(2)}\right). \label{quadoptwealth}
\end{align}

Apparently, the above two expressions of $X^{\tilde{u}^*}_T$ should match, and hence
\begin{equation} \label{quadtheta}
\begin{split}
\frac{1}{\alpha} \frac{\mu(t)-r(t)}{\sigma(t)} \frac{P_t}{P_T} Z_{t-} &= e^{r(T-t)} (\sigma(t)\tilde{\pi}^*(t)-\rho b L^*(t)), \\
\frac{1}{\alpha} \frac{P_t}{P_T} \theta_2(t) &= e^{r(T-t)} b \sqrt{1-\rho^2} L^*(t),\\
\frac{1}{\alpha} \frac{P_t}{P_T} \theta_3(t) &=e^{r(T-t)}  \gamma L^*(t).
\end{split}
\end{equation}

By \eqref{quadtheta}, we can rearrange \eqref{quadwealth} as
\begin{equation} \label{quadsystem}
\begin{split}
X^{\tilde{u}^*}_T &= \frac{1}{\alpha} - \frac{Z_0}{\alpha P_T} - \frac{1}{\alpha P_T} \int_0^T Z_{t-} \xi_t P_tdt  + \int_0^T e^{r(T-t)} \lambda \gamma L_t^* dt\\
&\quad + X^{\tilde{u}^*}_T - xe^{rT} - \int_0^T e^{r(T-t)} \big( (\mu_t-r_t)\tilde{\pi}^*_t+(p-a)L^*_t \big)dt.
\end{split}
\end{equation}

From the systems of \eqref{quadtheta1} and \eqref{quadtheta2} along with the above conditions \eqref{quadtheta}, we find optimal control as
\begin{align}
\tilde{\pi}^*(t)=e^{-r(T-t)} \frac{1}{\alpha} \frac{\mu(t)-r(t)}{\sigma(t)^2} \frac{P_t}{P_T} Z_{t-} +\frac{\rho b}{\sigma(t)} L^*(t), \label{quadpi}\\
L^*(t)=e^{-r(T-t)} \frac{1}{\alpha} \frac{p-a-\lambda \gamma +\rho b \frac{\mu(t)-r(t)}{\sigma(t)}}{b^2(1-\rho^2)+\lambda \gamma^2} \frac{P_t}{P_T} Z_{t-}, \label{quadkappa}
\end{align}

To ensure the equation \eqref{quadsystem} holds, we choose $\xi$ to be
\begin{equation} \label{quadxi}
\xi(t)=- \left(\frac{\mu(t)-r(t)}{\sigma(t)}\right)^2 - \varphi(t),
\end{equation}
with $\varphi$ defined by
\[\varphi(t):=\frac{ \left( p-a-\lambda \gamma +\rho b \frac{\mu(t)-r(t)}{\sigma(t)} \right)^2 }{b^2(1-\rho^2)+\lambda \gamma^2},\]
and $Z_0$ as
\begin{equation} \label{quadzeta_0}
Z_0=(1-\alpha e^{rT}) P_T=(1-\alpha e^{rT})\exp\left\{ \int_0^T \xi(t)dt \right\}.
\end{equation}

Now we substitute optimal $L^*$ into \eqref{quadtheta} and obtain the expressions of $\theta_2$ and $\theta_3$ in $Z$ as
\begin{align}
\theta_2(t) &= b\sqrt{1-\rho^2} \varphi(t) Z_{t-}, \label{quadtheta_2}\\
\theta_3(t) &= \gamma \varphi(t) Z_{t-}. \label{quadtheta_3}
\end{align}

Therefore, we obtain the dynamics of $Z$ as
\begin{equation*}
dZ_t=Z_{t-} \left( -\frac{\mu_t-r_t}{\sigma_t} dW_t^{(1)} + b \sqrt{1-\rho^2} \varphi_t dW_t^{(2)} + \gamma \varphi_t dM_t \right),
\end{equation*}
which yields a unique solution
\begin{equation} \label{quadzeta}
\begin{split}
Z_t =Z_0 \exp &\bigg\{ -\int_0^t \left( \frac{1}{2} \left( \frac{\mu_s-r_s}{\sigma_s} \right)^2 + \frac{1}{2} b^2 (1-\rho^2) \varphi^2_s + \lambda \gamma \varphi_s \right)ds \\
&\; \int_0^t \left( -\frac{\mu_s-r_s}{\sigma_s} dW_s^{(1)} + b \sqrt{1-\rho^2} \varphi_s dW_s^{(2)} + \gamma \varphi_s dN_s \right) \bigg\},
\end{split}
\end{equation}
where $Z_0$ is given by \eqref{quadzeta_0}.

Define $\Phi$ as
\[ \Phi(t):=\frac{p-a-\lambda \gamma +\rho b \frac{\mu(t)-r(t)}{\sigma(t)}}{b^2(1-\rho^2)+\lambda \gamma^2}.\]
Then we can rewrite optimal control in the following form
\begin{align}
\tilde{\pi}^*(t)&=\frac{1}{\alpha} e^{-r(T-t)} \left( \frac{\mu(t)-r(t)}{\sigma^2(t)} + \frac{\rho b}{\sigma(t)} \Phi(t) \right) \exp\left\{\int_t^T \xi(s)ds \right\} Z_{t-}, \label{quadpi1}\\
L^*(t) &= \frac{1}{\alpha} e^{-r(T-t)} \Phi(t) \exp\left\{\int_t^T \xi(s)ds \right\} Z_{t-}, \label{quadkappa1}
\end{align}
where $\xi$ and $Z$ are given by \eqref{quadxi} and \eqref{quadzeta}, respectively.

\begin{thm} \label{quadthm}
When $U(y)=y-\frac{\alpha}{2}y^2$, $\alpha>0$, and $\Phi(t)\ge0$, $\forall \, t\in[0,T]$, $\tilde{u}^*=(\tilde{\pi}^*,L^*)$ with $\tilde{\pi}^*$ and liabilities $L^*$
given by \eqref{quadpi1} and \eqref{quadkappa1}, respectively, is
optimal investment to Problem \ref{prob_opt} with the admissible control set $\mathcal{A}_2$.
\end{thm}
\emph{Proof.} $\forall \, t\in[0,T]$, $\Phi(t)$ is a bounded deterministic function, so are $\frac{\theta_i}{Z_{t-}}$, with $\theta_i$, $i=1,2,3$, defined by \eqref{quadtheta1}, \eqref{quadtheta_2} and \eqref{quadtheta_3}, respectively. Hence $Z$, defined by \eqref{quadzeta}, is indeed a square-integrable martingale. With our choices for $\xi$ and $Z_0$, given by \eqref{quadxi} and \eqref{quadzeta_0}, we can verify that
\begin{equation*}
- \frac{1}{\alpha P_T} \int_0^T Z_{t-} \xi_t P_tdt- \int_0^T e^{r(T-t)} \big( (\mu_t-r_t)\tilde{\pi}^*_t + (p-a- \lambda \gamma) L_t^* \big)dt=0, \\
\end{equation*}
and $\frac{1}{\alpha} - \frac{Z_0}{\alpha P_T}- xe^{rT} =0$, which implies $Z_T$ defined by \eqref{quadzeta} is equal to $1-\alpha X^{\tilde{u}^*}_T$, with $X^{\tilde{u}^*}_T$ given by \eqref{quadoptwealth}.

Provided $\tilde{u}\in \mathcal{A}_2$, $X^{\tilde{u}} \in L^2(\mathcal{F})$, and then $\tilde{Y} Z \in L^2(\mathcal{F})$, which verifies $\tilde{Y}Z$ is indeed a martingale under $\mathbb{P}$. So the condition \eqref{quadcond1} holds.

In the last step, we show that $\tilde{u}^*=(\tilde{\pi}^*,L^*)$, with $\tilde{\pi}^*$ and $L^*$ given by  \eqref{quadpi1} and \eqref{quadkappa1}, is admissible. To that purpose, notice both $\Phi(t)$ and $\xi(t)$ are bounded for all $t\in[0,T]$, so, $\forall \, t\in[0,T]$, there exists a positive constant $\tilde{K}_t$ such that
\[ \max\left\{ \left(\tilde{\pi}^*(t) \right)^2, \; \left(L^*(t) \right)^2 \right\} \le \tilde{K}_t \left(Z_{t-}\right)^2.\]
Due to the fact that $Z \in L^2(\mathcal{F})$, we obtain $\tilde{\pi}^*,\,L^*\in L^2(\mathcal{F})$. The assumption $\Phi(t)\ge 0$ guarantees that $L^*(t) \ge 0$, $\forall \, t \in[0,T]$. \hfill $\Box$

\section{Conclusions}
\label{sec_conclusion}
Motivated by the bailout case of AIG in the financial crisis and the increasing demand for efficient risk management in the insurance industry, we consider optimal investment and risk control problem for an insurer (like AIG). In our model, the insurer's risk is controllable and is assumed to follow a jump-diffusion process. As discussed in \citet[Chapter 6]{stein}, one major mistake AIG made is ignoring the negative correlation between its liabilities (risk) and the capital gains in the financial market. So we assume the risk process is negatively correlated with the performance of the financial market.

We consider a risk-averse insurer who wants to maximizer the expected utility of the terminal wealth by selecting optimal investment and risk control strategies.
We obtain explicit solutions to optimal strategies for logarithmic utility function, power utility function, exponential utility function and quadratic utility function.

\end{document}